\documentclass[aps,prc,onecolumn,superscriptaddress,showpacs,showkeys,nofootinbib]{revtex4-2}
\usepackage{graphicx}
\usepackage{bm}
\usepackage{epsfig}
\usepackage{amsfonts}
\usepackage{amssymb,amscd}
\usepackage{subfigure}
\usepackage{xcolor}
\usepackage{comment}
\usepackage{amsmath}

\begin{document}
%\begin{flushright}
%MS-TP-22-??
%\end{flushright}

\title{$D^0$ meson production in $pp$ collisions at large $Q_s^2$}

\author{Yuri N. {\sc Lima}}
\email{limayuri.91@gmail.com}
\affiliation{Institute of Physics and Mathematics, Federal %University of Pelotas, \\
  Postal Code 354,  96010-900, Pelotas, RS, Brazil}

\author{Andr\'e V. {\sc Giannini}}
\email{AndreGiannini@ufgd.edu.br}
\affiliation{
	Faculdade de Ci\^encias Exatas e Tecnologia, Universidade Federal da Grande Dourados (UFGD),
	Caixa Postal 364, CEP 79804-970 Dourados, MS, Brazil
}

\author{Victor P. {\sc Gon\c{c}alves}}
\email{barros@ufpel.edu.br}
\affiliation{Institute of Physics and Mathematics, Federal %University of Pelotas, \\
  Postal Code 354,  96010-900, Pelotas, RS, Brazil}
%\affiliation{Institut f\"ur Theoretische Physik, Westf\"alische %Wilhelms-Universit\"at M\"unster,
%Wilhelm-Klemm-Stra\ss e 9, D-48149 M\"unster, Germany}
\affiliation{Institute of Modern Physics, Chinese Academy of Sciences,
  Lanzhou 730000, China}

%----------------------------------------------------------------------
\begin{abstract}
The impact of the non-linear effects in the QCD dynamics on the observables is directly related to the magnitude of the saturation scale $Q_s$, which is predicted to increase with the energy, rapidity and multiplicity. In this paper, we investigate the $D^0$ meson production in $pp$ collisions at forward rapidities and/or high multiplicities considering the Color Glass Condensate (CGC) formalism and the solutions of the running coupling Balitsky - Kovchegov (BK) equation. The contributions of gluon -  and charm - initiated processes are taken into account, and a comparison with the current LHCb data is performed. The impact of an intrinsic charm component in the proton's wave function is also estimated. Predictions for   the self-normalized yields of $D^0$ mesons as a function of the multiplicity of coproduced charged hadrons are presented, considering $pp$ collisions at $\sqrt{s} = 13$ TeV and different values of the meson rapidity. A comparison with  the predictions for the kaon and isolated photon production is performed. Our results indicate that a future experimental analysis of the $D^0$ meson production at forward rapidities and high multiplicities can be useful to probe the CGC formalism and to disentangle the contribution of initial - and final - state effects.
\end{abstract}

\maketitle
\date{\today}

%----------------------------------------------------------------------
\section{Introduction}

Heavy quark production in hadronic collisions is considered one of the main tools for studying the properties of the
strong interactions~\cite{Andronic:2015wma}. The large quark mass $m_Q$ ensures the validity of perturbative calculations, and the strong dependence of the cross-section on the behavior of the gluon distribution makes its study ideal to probe the QCD dynamics at high energies.  One has that the huge density of low-$x$ gluons in the hadron wave-functions at high energies is expected to modify the usual description of the gluon distribution in terms of the linear DGLAP dynamics by the inclusion of non-linear corrections associated to the physical process of parton recombination \cite{Gelis:2010nm}.
This expectation can be easily understood: while for a large hard scale $Q^2$, the DGLAP equation predicts that the mechanism $g \rightarrow gg$ populates the transverse space with a large number of small size gluons per unit of rapidity, for small $Q^2$ the produced gluons overlap and fusion processes, $gg \rightarrow g$, are equally important \cite{Gribov:1983ivg}. The latter process implies that the rise of the gluon distribution below a typical scale is reduced, restoring the unitarity of the cross-sections. This scale is called saturation scale $Q_s$ and is predicted by the Color Glass Condensate (CGC) formalism \cite{CGC} to increase with the energy, atomic number, rapidity and multiplicity. In principle, for heavy quark production, the non-linear effects are expected to become negligible for $Q^2 \gg Q_s^2$, with $Q \propto \sqrt{p_T^2 + 4m_Q^2}$, where $p_T$ is the transverse momentum of the heavy state. In contrast, such effects are predicted to  modify the behavior of the cross-sections and differential distributions when the saturation scale becomes of the order or larger than $Q^2$ (See e.g. Refs.~\cite{Goncalves:2003ke,Kharzeev:2003sk,Fujii:2005vj}). 

In this paper, the heavy quark production in proton-proton ($pp$) collisions is studied at the LHC energies and in the regime where the inequality  $Q_s^2 \gtrsim Q^2$ is expected to be satisfied. For central rapidities, one has $Q_s^2 \approx 1$ GeV$^2$, which implies a negligible impact of non-linear effects. However, as $Q_s^2 \propto e^y$, these effects are enhanced for larger rapidities $y$ and are expected to modify the transverse momentum and rapidity distributions. In our analysis, we will compare our predictions, derived taking into account of the non-linear effects, with the LHCb data for the $D^0$ meson production at forward rapidities \cite{LHCb:20137TeV,LHCb:201613TeV}.
Moreover, we will also investigate the heavy quark production in high multiplicity events observed at the LHC. In the CGC formalism, high  multiplicity events are attributed to the presence of rare parton configurations (hot spots) in the hadrons that participate of the collision, which  are characterized by larger saturation scales in comparison to the typical configurations present in minimum bias events \cite{Ma:2018bax,Levin:2019fvb,Kopeliovich:2019phc,Gotsman:2020ubn,Siddikov:2020lnq,Siddikov:2021cgd,Stebel:2021bbn,Salazar:2021mpv,Lima:2022mol,Lima:2023dqw}. We will present predictions for the self-normalized yields of $D^0$ mesons as a function of the multiplicity of coproduced charged hadrons considering $pp$ collisions at $\sqrt{s} = 13$ TeV and different values of the meson rapidity. Moreover, our results will be compared with the predictions for the kaon and isolated photon production derived  in Refs.~\cite{Lima:2022mol,Lima:2023dqw}. As the impact of the  non-linear effects on these distinct final states is expected to be different, a future comparison of the predictions with the experimental data will be an important test of the CGC formalism, which predicts that the high multiplicity effects are only associated to initial - state effects, in contrast with the approaches proposed e.g. in Refs.~\cite{Ferreiro:2015gea,Werner:2016nsq,Tripathy:2022teu}.

\begin{figure}[t]
\begin{tabular}{ccc}
\includegraphics[scale=0.27]{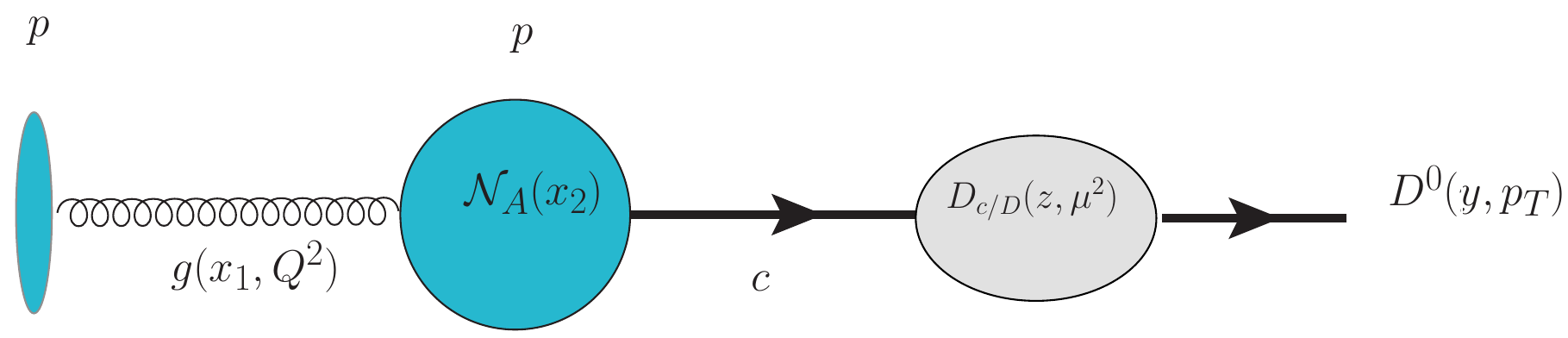} & \, & \includegraphics[scale=0.27]{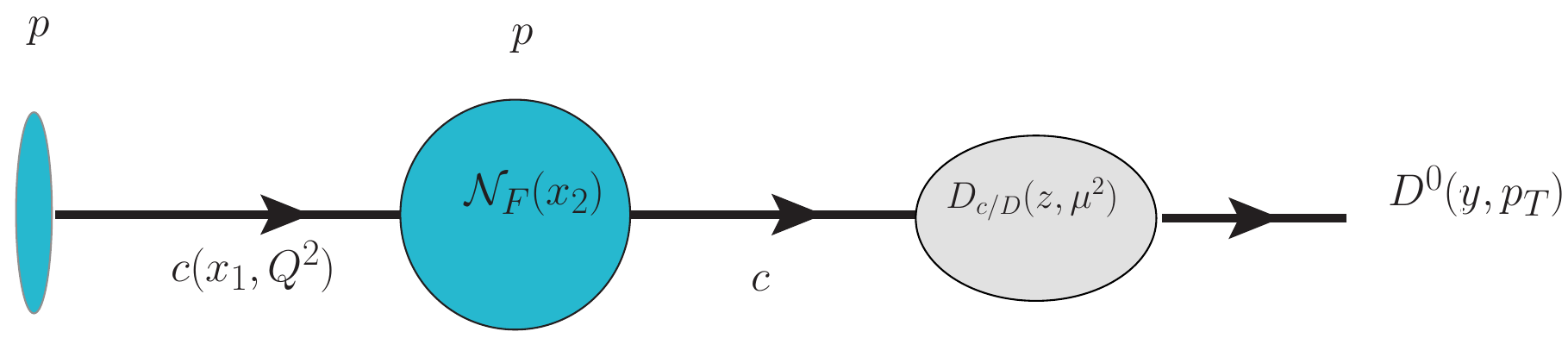}
\end{tabular}
\caption{Gluon (left)  and charm (right) - initiated contribution for the  production of a $D^0$ meson with transverse momentum $p_T$ and rapidity $y$ in $pp$ collisions.}
\label{Fig:diagram1}
\end{figure}

The $D^0$ meson production cross-section at forward rapidities 
will be estimated using the CGC formalism and taking into account of gluon and charm - initiated contributions, being represented schematically as follows (See Fig. \ref{Fig:diagram1})
\begin{eqnarray}
\sigma(pp \rightarrow D^0 X) \propto  g(x_1,Q^2) \otimes {\cal{N}_A}(x_2) \otimes D_{c/D} + c(x_1,Q^2) \otimes {\cal{N}_F}(x_2) \otimes D_{c/D} \,\,,
\label{Eq:sig_imp}
\end{eqnarray}
where $g(x_1,Q^2)$ and $c(x_1,Q^2)$ are the gluon and charm densities of the projectile proton, $D_{c/D}$ is the charm fragmentation function into the $D^0$ meson and the functions ${\cal{N}_A}(x_2)$ and ${\cal{N}_F}(x_2)$ are the adjoint and fundamental forward scattering amplitudes, which encodes all the information about the hadronic scattering, and thus about the non-linear and quantum effects in the hadron wave function. Such quantities will be estimated using the CGC formalism, by solving the running coupling Balitsky - Kovchegov (rcBK) equation \cite{BAL,KOVCHEGOV} for different initial conditions. 
The first term in Eq. (\ref{Eq:sig_imp}) can be associated to the $gg \rightarrow c \bar{c}$ subprocess and is expected to dominate at high center - of - mass energies and central rapidities. The charm initiated contribution is, in general, negligible in this same kinematical region. However, at forward rapidities, the $D^0$ meson production is dominated by collisions of projectile partons with large light cone momentum fractions ($x_1 \rightarrow 1$) with
target partons carrying a very small momentum fraction
($x_2 \ll 1$). Consequently, in addition to the small-$x$ effects in the target coming from the
non-linear aspects of QCD, large-$x$ effects in the projectile are also
expected to contribute. One of the possible new effects is the
presence of intrinsic heavy quarks in the hadron wave
function, which enhance the probability of finding  heavy quarks with large momentum fraction $x_1$ (For a review see, e.g. Ref.~\cite{Brodsky:2015fna}). The studies performed e.g. in Refs.~\cite{Goncalves:2008sw,Carvalho:2017zge,Giannini:2018utr,Maciula:2020dxv,Goncalves:2021yvw,Maciula:2022lzk}  have demonstrated that this contribution can become dominant at ultra-forward rapidities. Motivated by the recent evidences of an intrinsic charm (IC) component in the proton's wave function \cite{LHCb:2021stx,Ball:2022qks,NNPDF:2023tyk}, in this paper we revisit this topic and analyze the impact of an intrinsic charm on the transverse momentum and rapidity distributions for the $D^0$ meson production at the LHCb when the small-$x$ effects are described by the solution of the BK equation instead of  phenomenological models used in Refs.  ~\cite{Carvalho:2017zge,Goncalves:2017chx,SampaiodosSantos:2021tfh}. Moreover, we  investigate, for the first time,  its impact on the heavy quark production at high multiplicities.

This paper is organized as follows. The next section presents a brief review of the formalism used to estimate the $D^0$ meson production at forward rapidities together with the main ingredients in the calculations of the gluon and charm - initiated contributions. In particular, we discuss the different initial conditions of the BK equation used in our analysis, as well as the distinct models for the treatment of the intrinsic charm component. 
In Section \ref{Sec:Results} we present results for the transverse momentum and rapidity distributions associated to the $D^0$ meson production in $pp$ collisions at the LHC energies and compare with the LHCb data. Predictions for the Feynman-$x_F$ distribution and for the self-normalized yields of $D^0$ mesons as a function of the multiplicity of coproduced charged hadrons considering $pp$ collisions and different values of the meson rapidity are also presented. Finally, in Section \ref{Sec:conc}, we summarize our main results and conclusions.

\section{$D^0$ meson production in the forward region at high energies}
\label{section:formalism}

The heavy quark production in $pp$ collisions  is usually described considering the collinear factorization, where all partons involved are assumed to be on mass shell, carrying only longitudinal momenta, and their transverse momenta are neglected in the QCD matrix elements \cite{Collins:1991ty}. However, at high energies, the effects of the finite
transverse momenta of the incoming partons become important \cite{Catani:1990eg} and the non-linear QCD effects are predicted to contribute 
significantly \cite{Gelis:2010nm}. As a consequence, generalized factorization schemes should be considered to take into account of these effects. In recent years, several groups have discussed in detail such a subject (See, e.g. Refs~\cite{Nikolaev:2005qs,Gelis:2008rw,Kotko:2015ura}). In particular, at forward rapidities, where the produced particles originate from the scattering of a dilute projectile off a dense target, a hybrid factorization has been proposed and applied for the heavy quark production e.g. in Refs.~\cite{Raufeisen:2002ka,Tuchin:2004rb,Goncalves:2006ch,Cazaroto:2011qq,Tuchin:2012cd,Fujii:2013yja,Altinoluk:2015vax,Carvalho:2017zge,Goncalves:2017chx,SampaiodosSantos:2021tfh}. In this paper, we will consider the approaches discussed in Refs.~\cite{Carvalho:2017zge,Goncalves:2017chx}, which we refer for a detailed derivation of the expressions presented in this section.

Initially, let's discuss the gluon - initiated (G.I.) $D^0$ meson production, represented in the left panel of Fig. \ref{Fig:diagram1}. At forward rapidities and in the dipole frame, the process can be factorized in terms of the projectile gluon distribution $x_1g(x_1,Q^2)$, the fragmentation function $D_{c/D}$ and the heavy quark production via the gluon - proton scattering, which is described by the cross-section of the $g + p \rightarrow Q \bar{Q} X$ process. Such a quantity takes into account the fluctuation of the projectile gluon into a $Q \bar{Q}$ pair and the interaction of  a colorless three-body system
$gQ \bar{Q}$ with the color background field of the target proton. As demonstrated in Ref.~\cite{Goncalves:2017chx}, the differential distribution for the production of a $D^0$ meson with transverse momentum $p_T$ at rapidity $y$ can be expressed as follows:
\begin{eqnarray} \label{mes-CS}
	\left.\frac{d\sigma_{pp \rightarrow D^0 X}}{dy d^2p_T}\right|_{G.I.} = \int_{z_{\rm min}}^1 \frac{dz}{z^2} \,
	 x_1g(x_1,Q^2) \,\int_{\alpha_{\rm min}}^1 d\alpha \frac{d^3\sigma_{gp \to c\bar{c}X}}{d\alpha d^2 q_T }\,D_{c/D} (z,\mu^2) \,,
	 \label{eq:ppMX}
\end{eqnarray}
where $z$ is the fractional light-cone momentum of the charm $c$ carried by the meson, $\vec{q}_T = \vec{p}_T/z$,     $\alpha$ is the momentum fraction of the gluon carried by the charm   and  $D_{c/D}$ is the fragmentation function. Moreover, one has that
\begin{eqnarray} \label{Eq:limits}
z_{\rm min}=\frac{\sqrt{m_D^2+p_T^2}}{\sqrt{s}}\,e^{y}\,\,\,\,\mbox{and} \,\,\,\,
\alpha_{\rm min} = \frac{z_{\rm min}}{z}\sqrt{\frac{m_c^2 z^2 + p_T^2}{m_D^2 + p_T^2}}\,.
\end{eqnarray}
 The differential cross-section for the $g + p \rightarrow c \bar{c} X$ process is given by~\cite{Goncalves:2017chx}
\begin{eqnarray}
\frac{d^3\sigma_{gp \to c\bar{c}X}}{d\alpha d^2 q_T } &=& 
\frac{1}{ 6 \pi}  \int \frac{d^2 \kappa_T}{\kappa_T^4}  \alpha_s \, {\cal K}_{\rm dip}(x_2,\kappa_T^2)\,
\Big\{\Big[\frac{9}{8}{\cal{H}}_0(\alpha,\bar{\alpha},\vec{q}_T) - \frac{9}{4} {\cal{H}}_1(\alpha,\bar{\alpha},\vec{q}_T,\vec{\kappa}_T) \nonumber\\ 
&+& {\cal{H}}_2(\alpha,\bar{\alpha},\vec{q}_T,\vec{\kappa}_T) + \frac{1}{8}{\cal{H}}_3(\alpha,\bar{\alpha},\vec{q}_T,\vec{\kappa}_T)\Big] + 
\left[ \alpha \longleftrightarrow \bar{\alpha}\right]\Big\} \,,
\end{eqnarray}
with $\bar\alpha = 1 - \alpha$, and the auxiliary functions ${\cal{H}}_i$ defined by
\begin{eqnarray} \nonumber
{\cal{H}}_0(\alpha,\bar{\alpha},\vec{q}_T) & = & \frac{m_c^2 + (\alpha^2 + \bar \alpha^2)q_T^2}{(q_T^2 + m_c^2)^2} \,\,,\\
{\cal{H}}_1(\alpha,\bar{\alpha},\vec{q}_T,\vec{\kappa}_T)& = & \frac{m_c^2 + (\alpha^2 + \bar \alpha^2) \vec{q_T}\cdot 
(\vec{q_T} - \alpha \vec{\kappa}_T ) }{[(\vec{q_T} - \alpha \vec{\kappa}_T)^2 + m_c^2](q_T^2 + m_c^2)}
\,\,, \nonumber \\
{\cal{H}}_2(\alpha,\bar{\alpha},\vec{q}_T,\vec{\kappa}_T)& = & \frac{m_c^2 + (\alpha^2 + \bar \alpha^2) 
(\vec{q_T} - \alpha \vec{\kappa}_T )^2 }{[(\vec{q_T} - \alpha \vec{\kappa}_T)^2 + m_c^2]^2}
\,\,, \nonumber \\
{\cal{H}}_3(\alpha,\bar{\alpha},\vec{q}_T,\vec{\kappa}_T)& = & \frac{m_c^2 + (\alpha^2 + \bar \alpha^2) 
(\vec{q_T} + \alpha \vec{\kappa}_T )\cdot (\vec{q_T} - \bar{\alpha} \vec{\kappa}_T ) }
{[(\vec{q_T} + \alpha \vec{\kappa}_T)^2 + m_c^2][(\vec{q_T} - \bar{\alpha} \vec{\kappa}_T)^2 + m_c^2]}\,\,.
\end{eqnarray}
In addition, one has that $x_{1,2} = (M_{c\bar{c}}/\sqrt{s})\,e^{\pm y}$. As in Ref.~\cite{Goncalves:2017chx}, we will assume  that the hard scale $Q^2$ is equal to the square of the invariant mass of the $c\bar{c}$ pair ($M_{c\bar{c}} = 2\sqrt{m_c^2 + q_T^2}$). Moreover, following Ref.  \cite{Bhattacharya:2016jce}, we will assume $\alpha_s = 0.373$.  { The  main ingredient to estimate the $D^0$ meson differential cross-section is the dipole transverse momentum distribution (TMD)   ${\cal K}_{\rm dip}$,  which can be expressed in terms of the Fourier transform of the adjoint forward scattering amplitude ${\cal{N}_A}$ 
as follows
\begin{eqnarray}
{\cal K}_{\rm dip}(x,\kappa_T^2) = \frac{C_F}{(2\pi)^3} \int d^2\vec{r} e^{- i \vec{\kappa}_T \cdot \vec{r} } \,\,  \nabla_r^2 {\cal{N}_A}(x,r) \,\,,
\end{eqnarray}
and, consequently, is determined by the QCD dynamics at high energies. In particular, in the large - $N_c$ limit the adjoint scattering amplitude  can be obtained from the fundamental dipole scattering amplitude  ${\cal{N}_F}$, which is the solution of the rcBK equation, using the following relation: $ {\cal{N}_A}(x,r) = 2{\cal{N}_F}(x,r) - {\cal{N}}^2_F(x,r)$. 
In previous studies, ${\cal K}_{\rm dip}$ was estimated disregarding the non-linear effects and/or considering phenomenological models to describe them. In contrast, in our analysis, we will estimate such a quantity in terms of  solutions of the running coupling BK equation for three distinct initial conditions (see below).
}

For the charm - initiated (C.I.) contribution, represented in the right panel of Fig. \ref{Fig:diagram1}, we will consider the approach proposed in Ref.~\cite{Goncalves:2008sw} and rederived in the hybrid formalism in 
Ref.~\cite{Altinoluk:2015vax}. The basic idea is that a charm quark, present in the wave function of the incident proton, scatters off with the color background field of the target proton and then fragments into a $D^0$ meson. The hybrid formalism implies that the differential cross-section for the production of $D^0$ meson with transverse momentum $p_T$ at rapidity $y$ is given by~\cite{Goncalves:2008sw}
\begin{eqnarray}
	\left.\frac{d\sigma_{pp \rightarrow D^0 X}}{dy d^2p_T}\right|_{C.I.} = \frac{\sigma_0}{2(2\pi)^2}\int_{x_F}^{1}dx_1\frac{x_1}{x_F}\bigg[c(x_1,Q^2)\,\tilde{{\mathcal{N}}}_F\left(\frac{x_1}{x_F}p_T,x_2\right)D_{c/D}\left(z = \frac{x_F}{x_1},\mu^2\right)\bigg]\,\,,
\label{Eq:charm}	
\end{eqnarray}
where  $x_{1,2}$ represent the momentum fraction of the projectile and target parton that interact in the scattering process and $x_F$ is the Feynman-$x$ of the produced meson, which are defined by
\begin{equation}
	x_{1,2}=\frac{p_T}{z\sqrt{s}}e^{\pm y}
	\qquad\qquad \mbox{and} \qquad\qquad
	x_F = x_1 - x_2 \,\,.
\end{equation}
Moreover, $\tilde{{\mathcal{N}}}_F$ is the Fourier transform of the fundamental forward scattering amplitude, determined by solving the BK equation, and $\sigma_0$ is a constant obtained by  fitting the HERA data using the corresponding solutions (see below). { In our analysis, we will assume the values of $\sigma_0$ obtained in Ref. \cite{Albacete:2010sy}}.   As demonstrated e.g. in Refs.~~\cite{Goncalves:2008sw,Carvalho:2017zge,Giannini:2018utr,Maciula:2020dxv,Goncalves:2021yvw,Maciula:2022lzk}    such a contribution is negligible for central rapidities, but becomes important for forward rapidities and can dominate if an intrinsic component of charm is present in the proton's wave function. It is important to emphasize that recent results indicate the presence of this component ~\cite{LHCb:2021stx,Ball:2022qks,NNPDF:2023tyk}. Following Ref.~\cite{Carvalho:2017zge}, we will consider the models discussed in Ref.~\cite{Pumplin:2007wg}   to describe the intrinsic charm and a comparison with the results derived disregarding this component will also be provided.

\begin{figure}[t]
\includegraphics[scale=0.85]{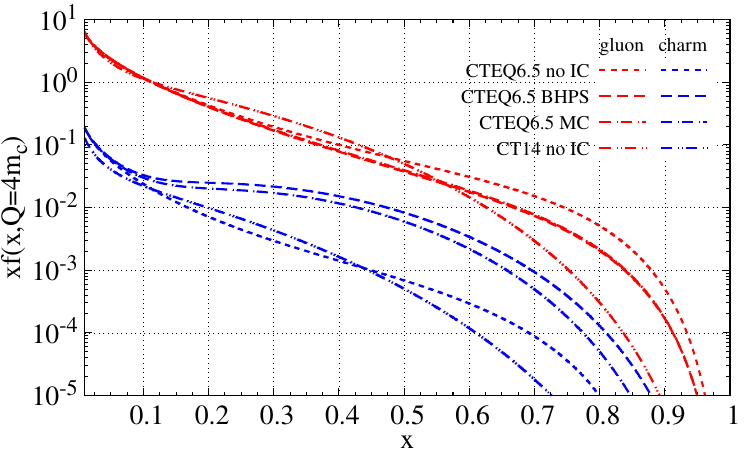} 
\caption{Predictions of the different intrinsic charm models for the $x$ - dependence of the charm (lower blue curves) and gluon (upper red curves) distributions, as obtained by the CTEQ 6.5 parametrization~\cite{Pumplin:2007wg}. For comparison, the results from CT14 parametrization~\cite{Dulat:2013hea} are also shown.}
\label{Fig:pdfs}
\end{figure}

\section{Results}
\label{Sec:Results}

In the next two subsections, we will present our results for prompt $D^0$ meson production at forward rapidities and at high multiplicities, respectively, considering the three solutions of the rcBK equation discussed in the previous section.  In order to estimate the $D^0$ meson differential cross-sections we will use the BKK05 parametrization~\cite{Kniehl:2006mw} to describe the fragmentation process. The collinear charm and gluon distributions will be described by the 
 CTEQ6.5~\cite{Pumplin:2007wg}, which provides a way of quantifying the impact of  an intrinsic charm component in the proton's wave function. A comparison with the predictions derived using the  CT14~\cite{Dulat:2013hea} set will also be provided. For completeness of our analysis, in  Fig. \ref{Fig:pdfs} we present a comparison between the  charm and gluon PDFs predicted by these different sets for a fixed hard scale $Q = 4m_c$; parameterizations disregarding an intrinsic charm component are denoted by ``no IC''. While in the CTEQ6.5 no IC and CT14 no IC sets the charm PDFs  are perturbatively generated by gluon splitting and comes from the DGLAP evolution, in the Brodsky - Hoyer - Peterson - Sakai (BHPS) \cite{Brodsky:1980pb} and Meson Cloud (MC) \cite{Navarra:1995rq,Paiva:1996dd,Hobbs:2013bia} models, a nonperturbative charm component is present in the proton's wave function. In the BHPS model, this component is associated to higher Fock states, as e.g. the $|uudc\bar{c}\rangle$ state. In contrast, in the MC model, it comes  e.g. from the nucleon fluctuation into an intermediate state
composed by a charmed baryon plus a charmed meson. One has that these models predict an  enhancement of the charm distribution at large $x$ $(> 0.1)$ (lower curves in Fig.~\ref{Fig:pdfs}) in comparison to the no IC predictions. Due to the momentum sum rule, the corresponding gluon distributions are also modified by the inclusion of
intrinsic charm, as observed in the upper curves of Fig.~\ref{Fig:pdfs}. As a consequence, the heavy quark production at large rapidities, which probes large values of $x$ in the projectile proton, is sensitive to the IC component. Such result has motivated the analysis performed in Refs.~\cite{Goncalves:2008sw,Carvalho:2017zge}, which will be improved in the present paper.

\begin{figure}[t]
\includegraphics[scale=0.85]{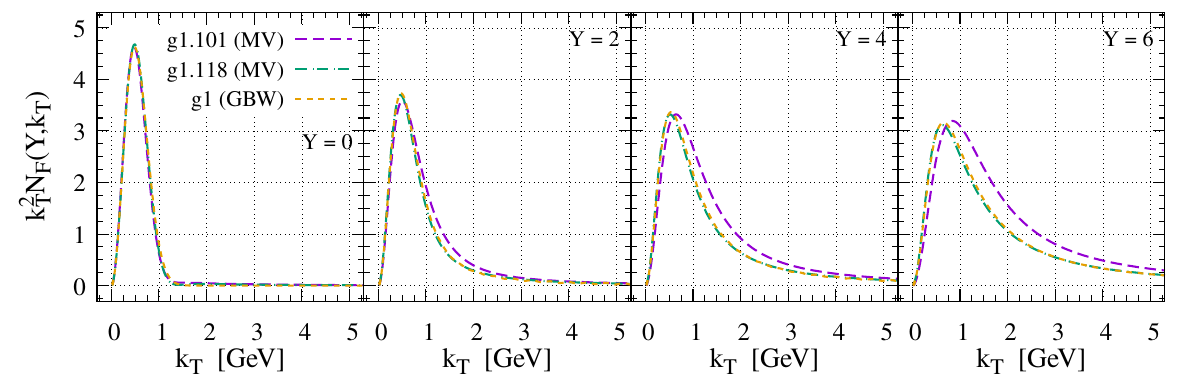} 
\caption{Comparison between the predictions for the Fourier transform of fundamental dipole scattering amplitude for different values of $Y = \ln(x_0/x)$, with $x_0 = 0.01$, derived considering different initial conditions for the running coupling BK equation.}
\label{Fig:ugds}
\end{figure}

 For the adjoint and fundamental forward dipole scattering amplitudes, needed to calculate ${\cal K}_{\rm dip}$ and $\tilde{{\mathcal{N}}}_F$, we will consider the solutions of the running coupling BK equation, which is the simplest non-linear evolution equation for the dipole-hadron scattering amplitude, being actually a mean field version of the first equation of the Balitsky hierarchy~\cite{BAL}. Such an equation is usually solved considering two distinct functional forms for its initial condition, inspired by the GBW~\cite{Golec-Biernat:1998zce} and  MV ~\cite{McLerran:1997fk} models,  given by  
\begin{gather}
{\mathcal{N}}(x_0 = 0.01,r) = 1 - \exp\bigg[-\frac{r^2 Q_{s,0}^2}{4}\bigg] \qquad\qquad\qquad\qquad \text{(GBW-like)}\\
{\mathcal{N}}(x_0 = 0.01,r) = 1 - \exp\bigg[-\frac{(r^2 Q_{s,0}^2)^{\gamma}}{4} \ln \left(\frac{1}{\Lambda r}+ e\right)\bigg] \qquad \text{(MV-like)}\,,
\end{gather}
where the free parameters, the initial saturation scale, $Q_{s,0}$ and the anomalous dimension, $\gamma$ are usually determined by fitting the HERA data. In our analysis, we will consider the parameters obtained in Refs.~\cite{Albacete:2009fh,Albacete:2010sy}. 
As in \cite{Lima:2023dqw} three distinct models are considered in order to investigate the role of the non-linear QCD dynamics: a GBW-like initial condition characterized by $Q_{s,0}^2 = 0.24 $ GeV$^2$ and two solutions assuming a MV-like initial condition where ($Q_0^2 = 0.157$ GeV$^2$, $\gamma = 1.101$) and ($Q_0^2 = 0.1597$ GeV$^2$, $\gamma = 1.118$).  Results obtained from the GBW-like initial condition are simply denoted by ``GBW'' and identify results with the MV-like initial condition by their value for the anomalous dimension $\gamma$: ``g1.101 (MV)'' and ``g1.118 (MV)''. In Fig. \ref{Fig:ugds} we present a comparison of the Fourier transform of the fundamental dipole scattering amplitude derived considering these different initial conditions. Results are shown as a function of the transverse momentum and fixed values of $Y = \ln(x_0/x)$, where $x_0 = 0.01$, which implies that for $Y=0$ we are presenting the initial conditions. One has that while all predictions are similar for small values of $Y$, differences start to appear for the g1.101 (MV) case when  $k_T\gtrsim 1$ GeV as one evolves in rapidity. Moreover, the peak of all distributions is shifted to larger values of $k_T$, which is directly associated to the increasing of the saturation scale at smaller values of $x$. Our results indicate that the predictions derived using  g1 (GBW) and g1.118 (MV) initial conditions are expected to be similar, differing from the g1.101 (MV) at large transverse momentum.

\begin{figure}[t]
\includegraphics[scale=0.7]{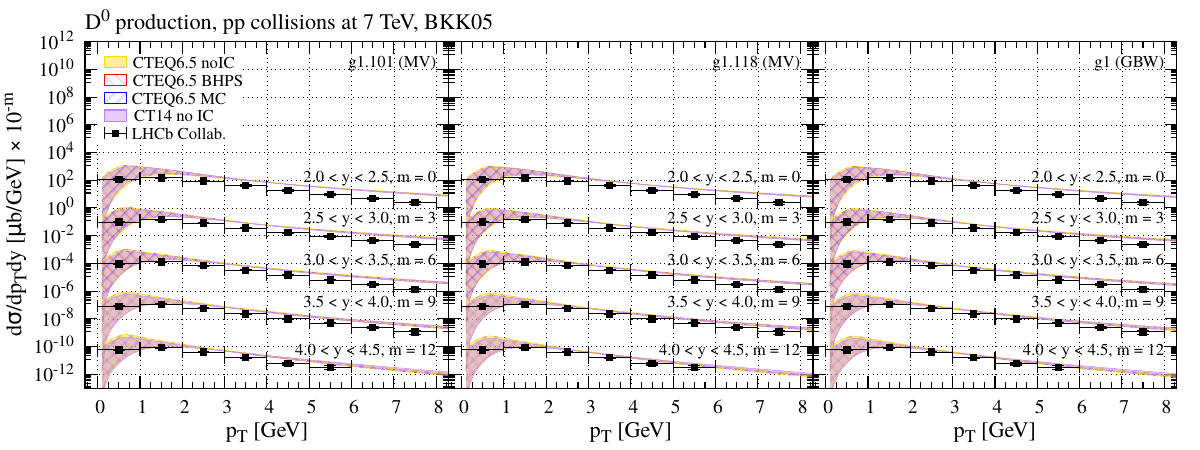}
\caption{Transverse momentum spectra of prompt neutral $D$ mesons for different rapidity bins at $\sqrt{s} = 7$ TeV. Results are presented for three different solutions of the rcBK equation, g1.101 (MV), g1.118 (MV), and g1 (GBW) and different models for the IC component. The experimental data is from Ref.~\cite{LHCb:20137TeV}.}
\label{Fig:7TeVpTspectrumD0production}
\end{figure}
%%%%
\begin{figure}[t]
\includegraphics[scale=0.7]{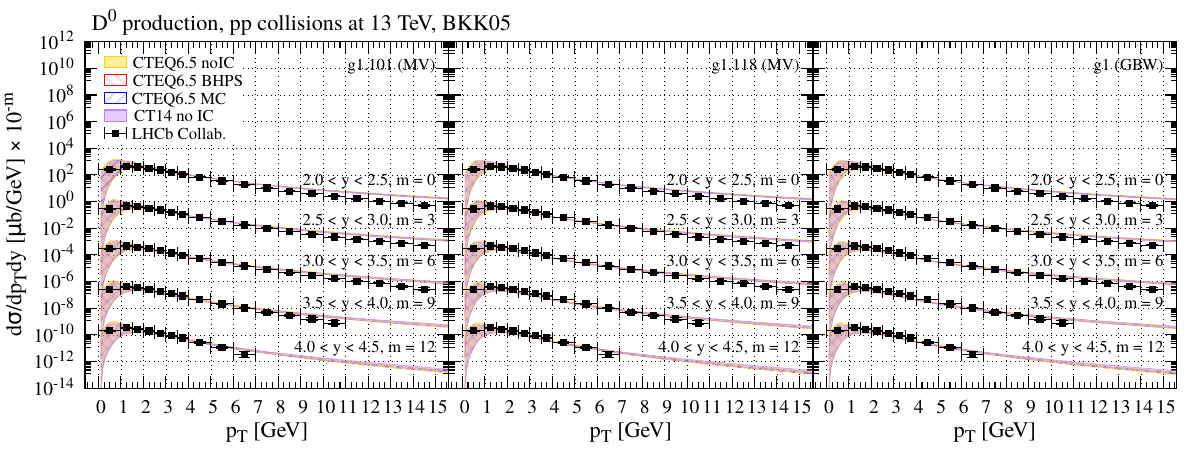}
\caption{Same as Fig. \ref{Fig:7TeVpTspectrumD0production} but for $\sqrt{s} = 13$ TeV. The experimental data is from Ref.~\cite{LHCb:201613TeV}.}
\label{Fig:13TeVpTspectrumD0production}
\end{figure}

\subsection{$D^0$ meson production at forward rapidities}

In Figs. \ref{Fig:7TeVpTspectrumD0production} and \ref{Fig:13TeVpTspectrumD0production} we present the $p_T$-spectra of $D^0$ mesons produced in $pp$ collisions at 7 and 13 TeV, respectively, calculated using the different solutions of the rcBK equation. Our results, which are calculated using the middle value of the appropriate rapidity bin, are compared to the LHCb data~\cite{LHCb:20137TeV,LHCb:201613TeV}; results were multiplied by a factor of $10^{-m}$ ($m = 0,2,4,6,8$) to improve visibility. The uncertainty band was generated by multiplying/dividing the momentum scale present in the PDFs and fragmentation function by a factor of two. One can see a good, overall agreement with experimental data is achieved, regardless of the BK solution or PDF considered. In particular, our results indicate that the current $D^0$ data for the $p_T$ spectra measured in the kinematical range probed by the LHCb detector is not sensitive to the intrinsic charm. 
One has that the error band increases at low $p_T$, which is associated with the limitations related to the description of the incoming projectile in the perturbative QCD framework. Results derived in~\cite{Goncalves:2017chx} indicate that the agreement may be improved by taking the transverse momentum of the projectile parton into account through a phenomenological model. A more careful treatment of the soft region may be achieved by including the Sudakov factor~\cite{Watanabe:2015yca,Benic:2022ixp}. Such inclusion, however, degrades the description of the data at higher $p_T$, which indicates that the treatment of the $D^0$ production at small - $p_T$ is still an open question.
In comparison with the predictions for $\sqrt{s} = 7$ TeV, the description of the experimental data for $\sqrt{s} = 13$ TeV becomes considerably better, specially for larger rapidities. It is important to emphasize that similar results have been obtained in calculations (including ones not necessarily based in the CGC formalism) performed by other groups~\cite{Goncalves:2017chx,Santos:2022eee,Xie:2021ycd,Bai:2021ira,SampaiodosSantos:2021tfh,Guiot:2021vnp,Maciula:2022lzk,Bhattacharya:2023zei}, indicating that the full description of these data still is  a challenge. Surely, the  forthcoming LHCb data from run 3 will be useful to clarify this aspect.

\begin{figure}[t]
\includegraphics[scale=0.7]{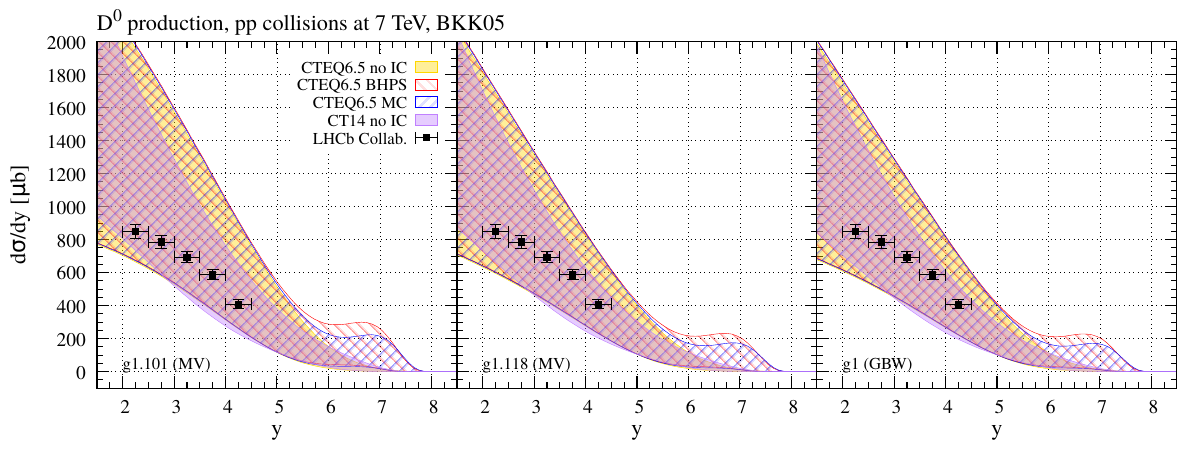}
\caption{Rapidity distribution of prompt neutral $D$ mesons produced in $pp$ collisions at $\sqrt{s} = 7$ TeV. Results are presented for three different solutions of the rcBK equation, g1.101 (MV), g1.118 (MV), and g1 (GBW) and different models for the IC component. The experimental data is from \cite{LHCb:20137TeV}.}
\label{Fig:7TeVyspectrumD0production}
\end{figure}

\begin{figure}[t]
\includegraphics[scale=0.7]{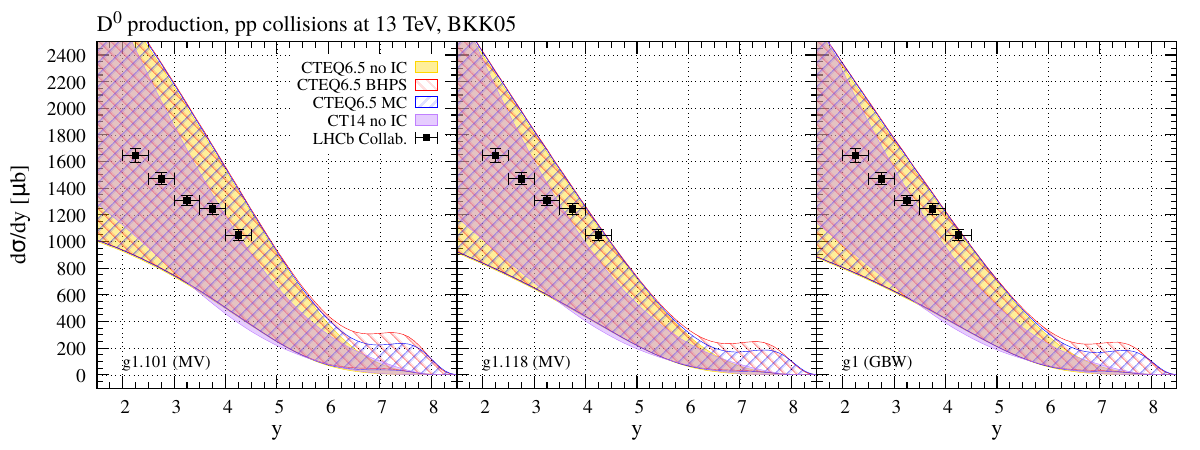}
\caption{Same as Fig. \ref{Fig:7TeVyspectrumD0production} but for $\sqrt{s}=13$ GeV. Data from Ref. \cite{LHCb:201613TeV}.}
\label{Fig:13TeVyspectrumD0production}
\end{figure}

Figs. \ref{Fig:7TeVyspectrumD0production} and \ref{Fig:13TeVyspectrumD0production} present our results for the rapidity distribution associated with the $D^0$ meson production in $pp$ collisions at 7 and 13 TeV, respectively. One has that the current experimental data can be described considering the different PDFs sets and the distinct solutions of the rcBK equation. The rather large uncertainty band can be traced back to the fact that $p_T$ integrated observables receive major contributions from soft, low momentum particles and this is exactly the regime where our calculations for the transverse momentum spectra present the largest uncertainties when varying the momentum scale in the PDFs and fragmentation functions. The reduction of the uncertainty band when moving to forward rapidities is purely kinematical and is related to the decrease of the available phase space; the slightly smaller uncertainty band for the calculation with the CT14 PDF may be linked to improvements made by including new data from Tevatron and LHC in their analysis~\cite{Dulat:2013hea}. Results that do not include an IC component, such as the CT14, decrease faster when moving to the ultra-forward region, while calculations including the intrinsic charm, present in the  CTEQ6.5 set, show a bump in a characteristic rapidity window. Such a bump moves to larger rapidities as the collision energy increases, since one needs to reach larger values of $y$ to enter the kinematic regime associated with the IC component.

%As in the $p_T$ spectra,  a better description of the rapidity distribution at 7 TeV is translated to the data for $d\sigma/dy$ now sitting roughly in the middle of our uncertainty band; at the same time, undershooting the $p_T$-spectra at 13 TeV leads to a rapidity distribution that is consistent with the experimental data in its upper limit.


%%% 

\begin{figure}[t]
\includegraphics[scale=0.7]{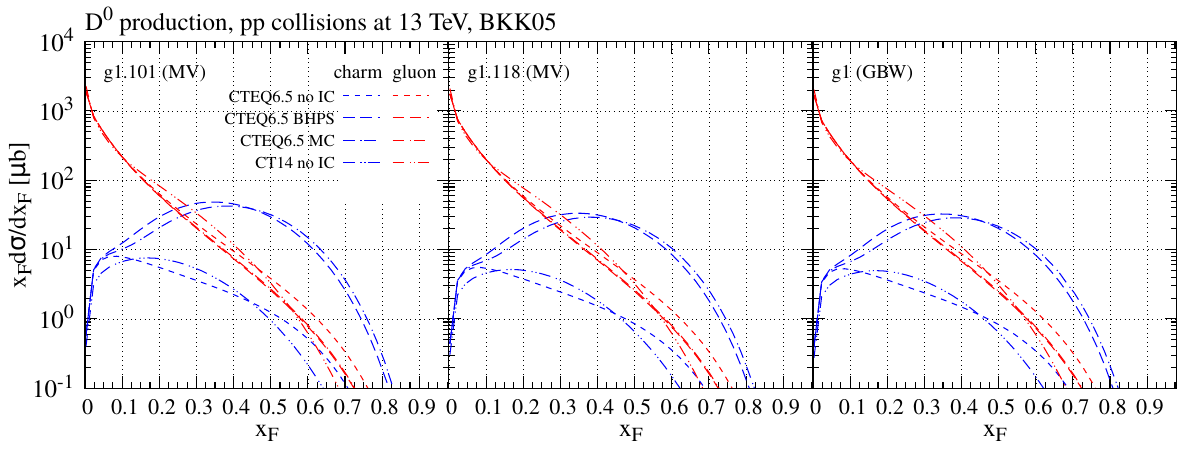}
\caption{Feynman-$x_F$ distribution associated to the $D^0$ meson production in $pp$ collisions at $\sqrt{s}=13$ TeV. Results are shown for the three solutions of the rcBK equation employed in this study.}
\label{Fig:13TeVxFspectrumD0prodcontribs}
\end{figure}

\begin{figure}[t]
	\includegraphics[scale=0.7]{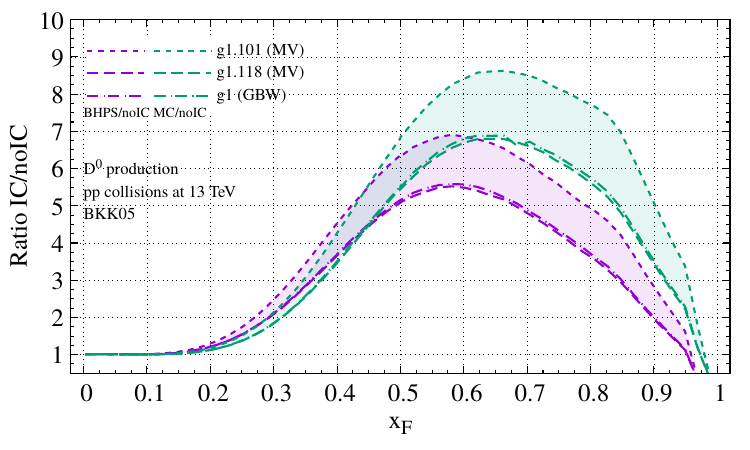}
	\caption{Ratio of Feynman-$x_F$ distributions with and without assuming an IC component  in the proton's wave function for the production of  $D^0$ mesons in $pp$ collisions at $\sqrt{s}=13$ TeV. }
	\label{Fig:Ratio_xFdistribution13TeV}
\end{figure}

The impact of an intrinsic charm component for the $D^0$ meson production can be better visualized in the  corresponding  Feynman-$x_F$ distributions, which can be obtained from the rapidity distributions using that 
\begin{gather}
x_F\frac{d\sigma_{pp\to D^0 X}}{dx_F} = \chi \frac{d\sigma_{pp\to D^0 X}}{dy}\,, \quad\qquad \text{where} \qquad
\begin{cases}
	\chi = 1, & {\text{for the charm-initiated channel}} \\
	\chi = \left(\sqrt{4M_{c\bar{c}}^2/x_F^2 s + 1}\right)^{-1}, & {\text{for the gluon-initiated channel}}\,.
\end{cases}
\end{gather}
In Fig. \ref{Fig:13TeVxFspectrumD0prodcontribs} we present the $x_F$ distribution for quark- and gluon-initiated channels, as well as, their sum, derived considering different solutions of the rcBK equation. One has that the results are almost independent of the solution used as input in the calculations. One can verify that the quark-initiated channel is always sub-dominant in the absence of an IC component. While such features are general and already shown, e.g. in Ref.~\cite{Carvalho:2017zge}, here we see that the splitting between the quark - and gluon - initiated channels are not as large as previously obtained with phenomenological models for the fundamental and adjoint dipole scattering amplitudes.  In contrast, the inclusion of an intrinsic component implies that the charm - initiated process dominates for $x_F \ge 0.3$, with its magnitude being dependent on the model assumed to describe the intrinsic charm, as demonstrated in Fig.~\ref{Fig:Ratio_xFdistribution13TeV}, where we present the ratio between the $x_F$ distributions estimated with and without the IC component. { The results indicate that the MC model implies a larger impact of the intrinsic charm on the $D^0$ meson production at large $x_F$.} The large uncertainty generated by the variation of the momentum scale appearing in the PDFs and fragmentation functions show the need of advancing in determining such ingredients with better precision. Progress regarding this matter is expected in the next decade, when the Electron-Ion Collider at BNL \cite{AbdulKhalek:2021gbh,Burkert:2022hjz} and Forward Physics Facility at LHC \cite{Anchordoqui:2021ghd,Feng:2022inv} are expected to enter in operation.

%Lastly, in Fig. \ref{Fig:Ratio_xFdistribution13TeV} we quantify the role of the IC component via a ratio of the Feynman-$x$ distributions. Even though the magnitude of this ratio is of the same order of the one presented in~\cite{Carvalho:2017zge}, the large uncertainty with the momentum scale appearing in the PDFs and fragmentation functions show the need of advancing in determining such ingredients with better precision. Progress regarding this matter is expected in the next decade when the Electron-Ion Collider is expected to enter in operation {\color{red}\bf [CITE]}. 

\subsection{$D^0$ meson production in high multiplicities events}
Over the last years, different experimental collaborations at the Relativistic Heavy Ion Collider (RHIC) and at the Large Hadron Collider (LHC) have studied the particle production in high multiplicities events at $pp$ collisions \cite{ALICE:2015ikl,ALICE:2017wet,STAR:2018smh,ALICE:2020msa,ALICE:2020eji,ALICECollaboration:2020, ALICE:2021zkd} and found, in particular, that the $D$ meson yield, measured at central rapidities,  grow rapidly as a function of the multiplicities of coproduced charged particles. Currently, these data can be described by
models based on very distinct underlying assumptions and physical mechanisms, and it is not clear if  
 the modification observed in high multiplicity events compared to the minimum bias case is due
to either initial - or final - state effects or both. Surely, future data, also for forward rapidities, will be useful to discriminate between these different models. Here, we investigate the dependence on $y$ of the enhancement in the $D^0$ meson production for a model based on the CGC formalism.  As in  Refs.~\cite{Ma:2018bax,Levin:2019fvb,Kopeliovich:2019phc,Gotsman:2020ubn,Siddikov:2020lnq,Siddikov:2021cgd,Stebel:2021bbn,Salazar:2021mpv,Lima:2022mol,Lima:2023dqw} we based our analysis in  two main assumptions: $i)$ we assume that the $D^0$ meson production, as discussed in the previous section, to also be valid when studying high-multiplicity events and $ii)$ collisions that generate more $D$ mesons are associated with interactions where the target is in a configuration with a larger saturation scale. One has that in the CGC formalism, the enhancement is associated to rare configurations in the proton wave function and, therefore, is associated to an initial - state effect, and has its rapidity dependence fully determined by the $x$ dependence of $Q_s$. As a consequence, a future experimental analysis of the $D^0$ meson production in high multiplicities events at forward rapidities will be a test of the CGC formalism and useful to clarify if final state effects should be included in the description of these events. { It is important to emphasize that the correlation between the normalized $D^0$ meson and charged particles yields in $pp$ collisions at $\sqrt{s} =$ 7 TeV was previously discussed in Ref. \cite{Ma:2018bax} and a comparison with the ALICE data for $y = 0$ was performed. In that study, they have used the formalism derived in Ref. \cite{Blaizot:2004wv}, which takes into account the contribution of the 2-, 3- and 4 - point correlators of Wilson lines. Moreover, they also included the contribution of  non-linear effects in the description of the projectile. The  contribution of the charm - initiated channel was not taken into account. In contrast, in the formalism considered in this paper, the differential cross-section for the gluon - initiated contribution is fully determined by the dipole TMD ${\cal K}_{\rm dip}$ in the transverse momentum space or, in an equivalent way, in terms of the dipole - proton cross-section in the impact parameter space. Such a quantity can be expressed in terms of the 2 - point correlator. Therefore, in our analysis, the contributions of the higher point correlators are not taken into account. Currently, the magnitude of these correlators in the LHC kinematical range is still a subject of intense study (See, e.g. Refs. \cite{Alvioli:2012ba,Marquet:2016cgx,Shi:2017gcq,Fukushima:2017mko,Banu:2019uei}).  In addition, non-linear effects in the projectile are not included in our calculations, which implies that our predictions are expected to be valid for forward rapidities, where the projectile wave function is probed at large - $x$. Certainly, a future comparison between the formalisms can be useful to improve our understanding of the heavy quark production at high partonic densities. We postpone such analysis for a forthcoming study. Regarding the current ALICE data for $y = 0$, we have verified that our predictions describe the data for small multiplicities, but fail for the higher values of normalized charged particle yield. Such result is expected, since for high multiplicities and central rapidities, we expect a non-negligible contribution of non-linear effects in the projectile, which are not included in our calculations.}

\begin{figure}[t]
 \begin{tabular}{ccc}
\includegraphics[scale=0.47]{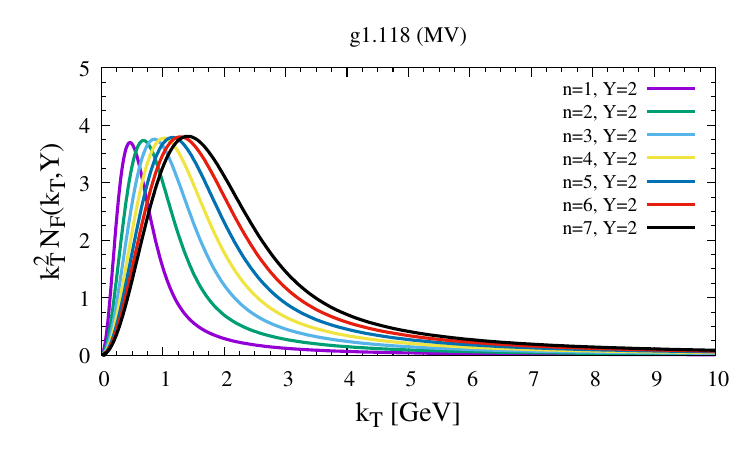} &
\includegraphics[scale=0.47]{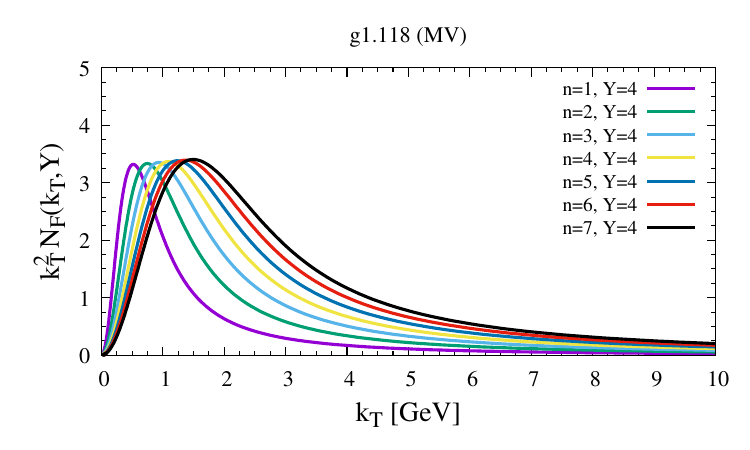} &
\includegraphics[scale=0.47]{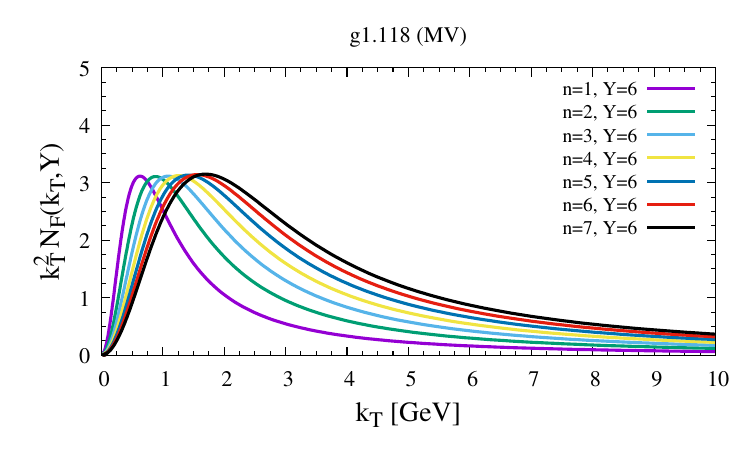}\
\end{tabular}
\caption{Dependence on the multiplicity $n$ of the Fourier transform of fundamental forward scattering amplitude. Results presented for distinct values of $Y = \ln (x_0/x)$.}
\label{Fig:enesdifene}
\end{figure}

The basic idea in the CGC formalism is that the hadron wave functions can be characterized by the saturation scale, which fluctuates event - by - event. In this framework, high multiplicity events are attributed to the presence of rare parton configurations (hot spots) in the hadrons that participate of the collision. Such highly occupied gluon states have larger saturation scales in comparison to the typical configurations present in minimum bias events.
These high multiplicity configurations can be approximated by increasing the value of $Q_s$ as follows: $Q_s^2(x,n) = n \cdot Q_s^2(x)$, with $n$ characterizing the increasing in multiplicity with respect to the minimum bias case.
As in Refs.~\cite{Ma:2018bax,Siddikov:2020lnq,Siddikov:2021cgd,Stebel:2021bbn,Salazar:2021mpv,Lima:2022mol,Lima:2023dqw}, such an increasing in the saturation scale will be taken into account by rescaling the initial condition for the rcBK equation, i.e. the saturation scale for an event with a multiplicity $n$ is obtained by solving the rcBK equation for an initial saturation scale given by $Q_{s,0}^2(n) = n \cdot Q_{s,0}^2$, with $Q_{s,0}^2$ determined by fitting the HERA data, as discussed in the previous subsection. Fig. \ref{Fig:enesdifene} illustrates the impact of this shift on the Fourier transform of fundamental forward scattering amplitude for different values of $Y = \ln (x_0/x)$ derived by solving the rcBK equation for distinct values of $n$. One has that the increasing of $n$ implies that the peak is displaced for larger values of the transverse momentum $k_T$, which is expected due to the increasing of the saturation scale. Moreover, a longer tail in $k_T$  is generated. A similar behavior is also verified in the Fourier transform of adjoint forward dipole scattering amplitude. As a consequence, the increasing with  the multiplicity of the  $D^0$ meson production is expected to be rapidity dependent. 

\begin{figure}[t]
\includegraphics[scale=0.8]{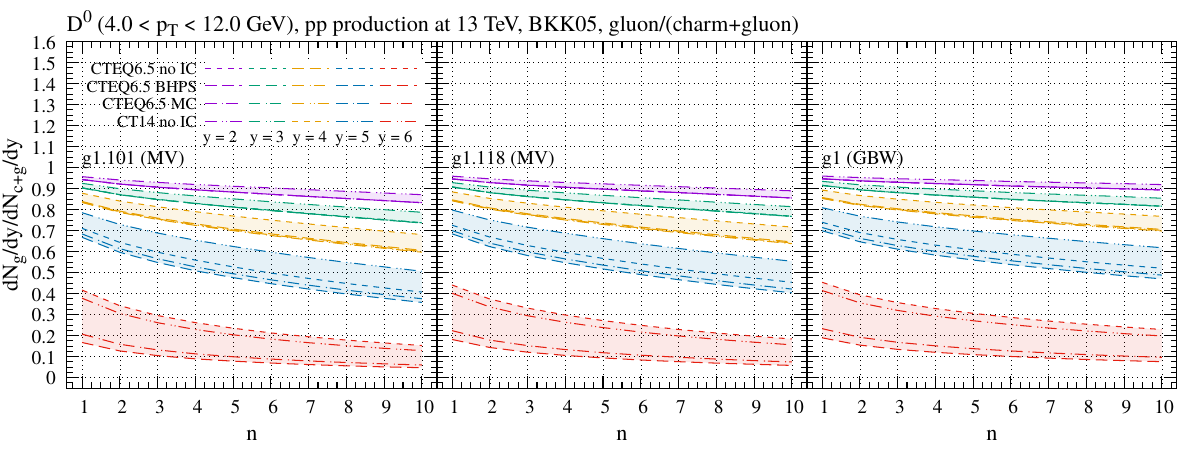}
\includegraphics[scale=0.8]{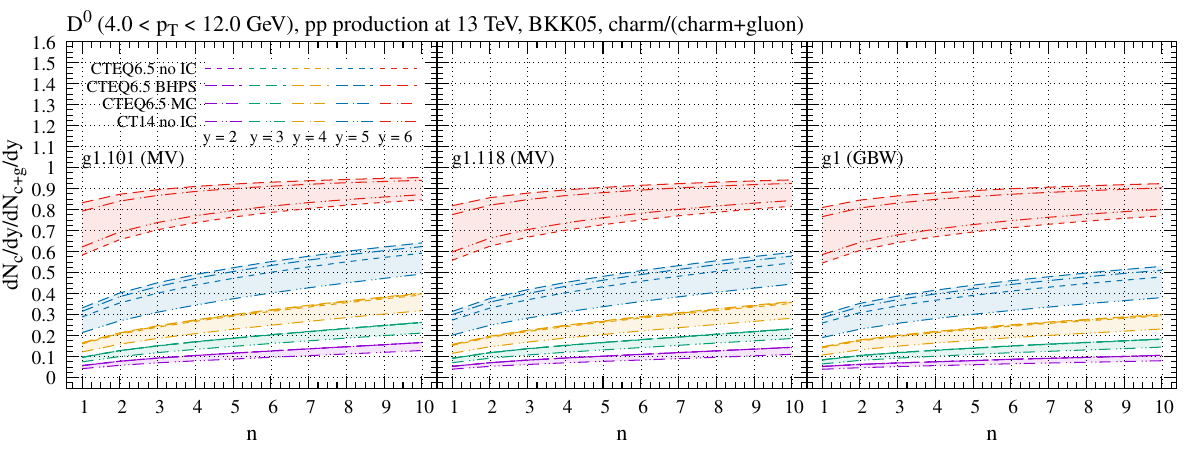}
\caption{Dependence on the scale factor $n$ of the ratios $dN_g/dy / (dN_{c+g}/dy)$  (upper panels) and $dN_c/dy / (dN_{c+g}/dy)$  (lower panels)   for different rapidities, PDFs and distinct solutions of the rcBK equation. }
\label{Fig:ndependence}
\end{figure}

{ Before present our predictions for the $D^0$ - meson production at high multiplicities, it is interesting to investigate the  dependence on the scale factor $n$ of the gluon  and charm - initiated channels for distinct rapidities. As discussed in the previous section, the contribution of charm - initiated channel increases with the rapidity and is expected to become dominant at ultra - forward rapidities. Such an expectation is observed in the results presented in Fig. \ref{Fig:ndependence}, where we show the predictions for the ratio between a given channel and the sum of them. The results have been derived integrating the $D^0$ transverse momentum in the range $4.0  < p_T < 12.0$ GeV and considering different PDFs and distinct solutions of the rcBK equation. The results are similar for distinct rcBK solutions, but dependent on the PDF used as input of calculations. Such result is expected, since the magnitude of the charm - initiated channel is strongly dependent on the presence (or not) of an intrinsic charm. Moreover, our results indicate that for $y \approx 5$, the contributions of gluon and charm - initiated channels are similar for the $D^0$ - meson production at high multiplicities (large $n$). }

\begin{figure}[t]
\includegraphics[scale=0.9]{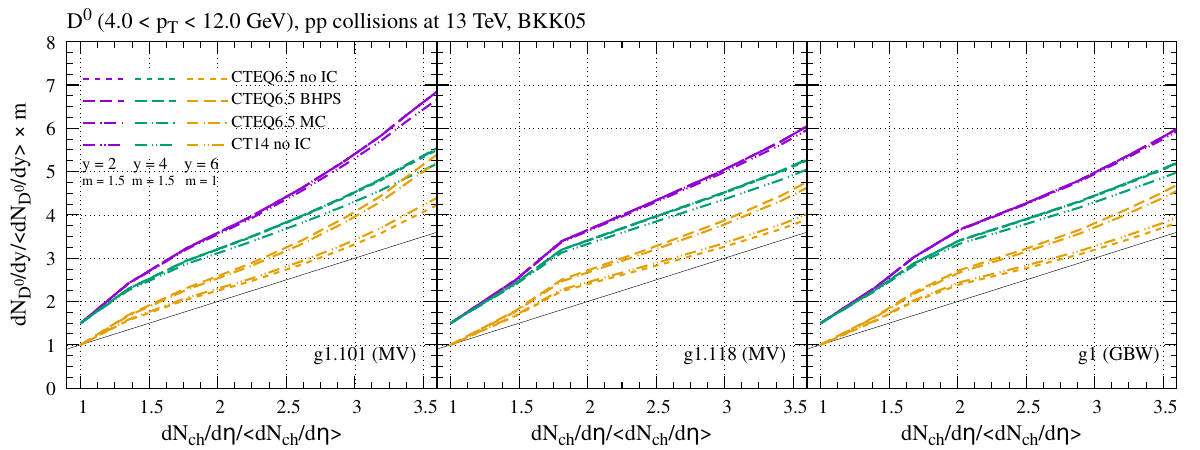}
\caption{Correlation between the normalized $D^0$ meson and charged particles yields in $pp$ collisions at $\sqrt{s} =$ 13 TeV, for three distinct solutions of the BK equation, different values of rapidity ($y= 2, 4, 6$) and comparing four different PDFs. The results were multiplied by a constant factor in order to improve visibility.}
\label{Fig:Dmeson_vs_ChParticles_PDFs}
\end{figure}

In Fig.~\ref{Fig:Dmeson_vs_ChParticles_PDFs} we present our predictions  for the 
correlation between the normalized $D^0$ meson and charged particles yields in $pp$ collisions at $\sqrt{s} =$ 13 TeV for  different values of the $D^0$ meson rapidity ($y= 2, 4, 6$). The results were multiplied by a constant factor in order to improve visibility. The predictions have been derived considering three distinct solutions of the BK equation and four different PDFs. The normalized yield for charged particles is  computed { using the $k_T$ - factorization formalism as detailed in Appendix C of Ref. ~\cite{Ma:2018bax}, which we refer for details,} integrating the differential distribution in the kinematical pseudorapidity range $|\eta| < 0.5$ and all possible values of the transverse momentum of the charged particles\footnote{{ We have verified that the  predictions derived using the distinct solutions of rcBK equation describe the current data for the inclusive hadron production at central rapidities in $pp$ collisions at the LHC.}}, which implies that this quantity is strongly modified by the saturation effects. Results for $D^0$ mesons, however, are for the fixed $4.0 \le p_T\, ({\rm GeV} ) \le 12$ range, as usually considered by experimental collaborations; predictions for other $p_T$ ranges can be provided upon request. { One has that the results for $y = 2$ and 4 are almost identical for distinct PDFs. However, for $y = 6$, the impact of an IC component becomes visible, with the IC predictions increasing faster with the multiplicity, which is expected from the results presented in Fig. \ref{Fig:ndependence}}. In addition, the predictions are sensitive to the solution of the rcBK equation considered in the calculation, which is also expected since in the CGC picture the increasing with the multiplicity is determined by the saturation scale that is determined by the QCD evolution. Finally, our results indicate that the enhancement is dependent on the rapidity. In particular, we have verified that for ultra - forward rapidities  one has $dN_{D^0}/dy/\langle dN_{D^0}/dy \rangle \approx dN_{ch}/d\eta/\langle dN_{ch}/d\eta \rangle$.  This behavior is associated with the increasing of the magnitude of saturation effects for larger rapidities, which implies that the impact on the $D^0$ meson production with $4.0 \le p_T \le 12$ GeV   become similar to that present in the charged hadron production. Such equality occurs for smaller rapidities if $D^0$ mesons with smaller $p_T$ are considered. Such aspects are a direct probe of the CGC formalism for high multiplicity events.

\begin{figure}[t]
\includegraphics[scale=0.9]{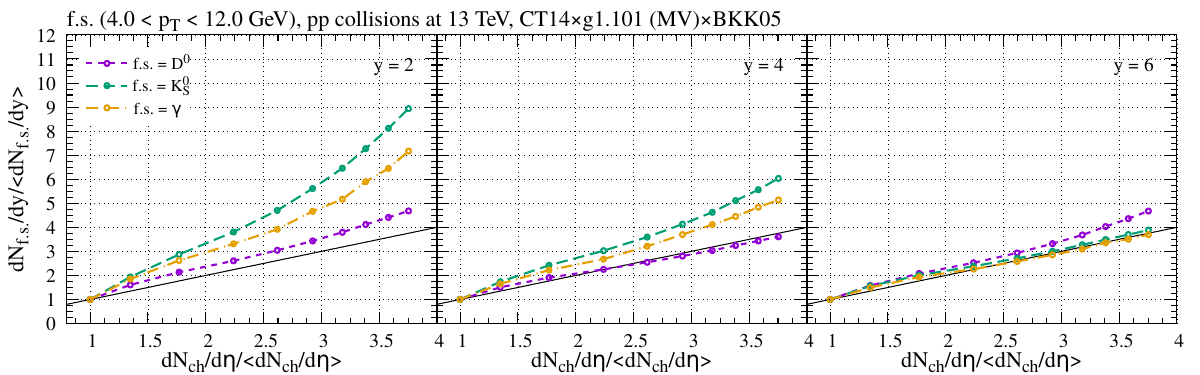}
\caption{Correlation between the normalized neutral $D$ mesons, $K^0_S$ and photons yields with charged particles for different values of rapidity.}
\label{Fig:D0K0gamma_vs_ChParticles}
\end{figure}

Finally, in Fig.~\ref{Fig:D0K0gamma_vs_ChParticles} we present a comparison between the predictions for the correlation between the normalized $D^0$ mesons with charged particles for different values of rapidity with the results for $K^0_S$ and photons yields derived in Refs.~\cite{Lima:2022mol,Lima:2023dqw}  also using the CGC formalism. Such an analysis verifies the general expectation of the CGC picture: if the high multiplicity events are solely a product of initial - state effects, the observables considered here should tend to a universal, final - state independent, curve at (ultra-)forward rapidities. In particular, isolated photons are not expected to be strongly enhanced when compared to massive final - states. If the enhancement is associated to final - state effects, though, such observables are expected to be modified in distinct ways for each particle specie. In contrast, in the CGC formalism, which provides a unified description of these processes, the presence of rare configurations with larger saturation scales will affect all observables, with the magnitude being dependent on the relation between $Q_s$ and the hard scale present in the process, which is given by a combination between the transverse momentum and the mass of the final state considered. Our results indicate that the slope is smaller for heavier states and increases for very forward rapidities, which is directly associated to the contribution of the charm - initiated channel. The comparison of these predictions with future experimental data can help us to disentangle initial - and final - state effects, as well as allow us to check the validity of the CGC formalism.

%----------------------------------------------------------------------

\section{Summary}
\label{Sec:conc}
In this paper we have investigated the $D^0$ meson production in $pp$ collisions at forward rapidities and/or high multiplicities, which are the regimes where we expect larger values of the saturation scale and, consequently, a higher impact of the non-linear effects of QCD dynamics on the differential distributions. As the contribution of the large - $x$ components of the proton's wave function also become important to describe the cross-sections in this regime, we also have analyzed the impact of an intrinsic charm component in the transverse and rapidity distributions. The process was described using the CGC formalism, and three distinct solutions of the rcBK equation were considered. We have demonstrated that the current LHCb data are well described and that the impact of the IC component is small. However, for higher rapidities, the charm - initiated process becomes dominant and is strongly sensitive  to the description of the charm PDF. We have also presented predictions for the self-normalized yields of $D^0$ mesons as a function of the multiplicity of coproduced charged hadrons considering different values of the meson rapidity. Our results indicated that the enhancement at high multiplicities is dependent on the rapidity, which is directly associated in the CGC formalism to the behavior of the saturation scale. A comparison with the predictions for other specific final states (kaon and isolated photon), also derived using the CGC formalism, have been performed. Our results indicate that a future experimental analysis of these three observables at forward rapidities and high multiplicities can be useful to probe the CGC formalism and to disentangle the contribution of initial and final-state effects.

\section*{Acknowledgements}
V.P.G. thanks A. Andronic for fruitful discussions about the measurement of high multiplicity events at the LHC.
This work was partially supported by INCT-FNA (Process No. 464898/2014-5).
V.P.G. was partially supported by CNPq, CAPES and FAPERGS. Y.N.L. was partially financed by CAPES (process 001). 
A.V.G. has been partially supported by CNPq. The authors acknowledge the National Laboratory for Scientific Computing (LNCC/MCTI, Brazil), through the ambassador program (UFGD), subproject FCNAE for providing HPC resources of the SDumont supercomputer, which have contributed to the research results reported within this study. URL: \url{http://sdumont.lncc.br}

%----------------------------------------------------------------------

%----------------------------------------------------------------------

\begin{thebibliography}{99}

\bibitem{Andronic:2015wma}
A.~Andronic, F.~Arleo, R.~Arnaldi, A.~Beraudo, E.~Bruna, D.~Caffarri, Z.~C.~del Valle, J.~G.~Contreras, T.~Dahms and A.~Dainese, \textit{et al.}
%``Heavy-flavour and quarkonium production in the LHC era: from proton\textendash{}proton to heavy-ion collisions,''
Eur. Phys. J. C \textbf{76}, no.3, 107 (2016)


\bibitem{Gelis:2010nm}
F.~Gelis, E.~Iancu, J.~Jalilian-Marian and R.~Venugopalan,
%``The Color Glass Condensate,''
Ann. Rev. Nucl. Part. Sci. \textbf{60}, 463-489 (2010)


\bibitem{Gribov:1983ivg}
L.~V.~Gribov, E.~M.~Levin and M.~G.~Ryskin,
%``Semihard Processes in QCD,''
Phys. Rept. \textbf{100}, 1-150 (1983)


\bibitem{CGC}  
J. Jalilian-Marian, A. Kovner, L. McLerran, and H. Weigert, Phys. Rev. D {\bf 55}, 5414 (1997);\\
J. Jalilian-Marian, A. Kovner, and  H.
Weigert, Phys. Rev. D {\bf 59}, 014014 (1999), {\it ibid.} {\bf 59}, 014015 (1999),
{\it ibid.} {\bf 59}  034007 (1999);\\
A. Kovner, J. Guilherme Milhano, and  H. Weigert, Phys. Rev. D {\bf 62}, 114005 (2000);\\
H. Weigert, Nucl. Phys. {\bf A703}, 823 (2002);\\
E. Iancu, A. Leonidov, and L. McLerran,
Nucl.Phys. {\bf A692}, 583 (2001);\\
E. Ferreiro, E. Iancu, A. Leonidov, and L. McLerran,
Nucl. Phys. {\bf A701}, 489 (2002).




\bibitem{Goncalves:2003ke}
V.~P.~Goncalves and M.~V.~T.~Machado,
%``Geometric scaling in inclusive charm production,''
Phys. Rev. Lett. \textbf{91}, 202002 (2003)



\bibitem{Kharzeev:2003sk}
D.~Kharzeev and K.~Tuchin,
%``Open charm production in heavy ion collisions and the color glass condensate,''
Nucl. Phys. A \textbf{735}, 248-266 (2004)

\bibitem{Fujii:2005vj}
H.~Fujii, F.~Gelis and R.~Venugopalan,
%``Quantitative study of the violation of k-perpendicular-factorization in hadroproduction of quarks at collider energies,''
Phys. Rev. Lett. \textbf{95}, 162002 (2005)



\bibitem{LHCb:20137TeV}
R. Aaij \textit{et al.} [LHCb Collaboration],
%``Prompt charm production in pp collisions at $\sqrt{s}=7$ TeV,''
Nucl. Phys. {\bf B871}, 1 (2013).

\bibitem{LHCb:201613TeV}
R. Aaij \textit{et al.} [LHCb Collaboration],
%``Measurements of prompt charm production cross-sections in pp collisions at $\sqrt{s}=13$ TeV,''
JHEP {\bf 03}, 159 (2016).


\bibitem{Ma:2018bax}
Y.~Q.~Ma, P.~Tribedy, R.~Venugopalan and K.~Watanabe,
%``Event engineering studies for heavy flavor production and hadronization in high multiplicity hadron-hadron and hadron-nucleus collisions,''
Phys. Rev. D \textbf{98}, no.7, 074025 (2018)





\bibitem{Levin:2019fvb}
E.~Levin, I.~Schmidt and M.~Siddikov,
%``Multiplicity dependence of quarkonia production in the CGC approach,''
Eur. Phys. J. C \textbf{80}, no.6, 560 (2020)


\bibitem{Kopeliovich:2019phc}
B.~Z.~Kopeliovich, H.~J.~Pirner, I.~K.~Potashnikova, K.~Reygers and I.~Schmidt,
%``Heavy quarkonium in the saturated environment of high-multiplicity $pp$ collisions,''
Phys. Rev. D \textbf{101}, no.5, 054023 (2020)


\bibitem{Gotsman:2020ubn}
E.~Gotsman and E.~Levin,
%``High energy QCD: multiplicity dependence of quarkonia production,''
Eur. Phys. J. C \textbf{81}, no.2, 99 (2021)



\bibitem{Siddikov:2020lnq}
M.~Siddikov and I.~Schmidt,
%``Multiplicity dependence of \ensuremath{\chi}c and \ensuremath{\chi}b meson production,''
Phys. Rev. D \textbf{104}, no.1, 016023 (2021)


\bibitem{Siddikov:2021cgd}
M.~Siddikov and I.~Schmidt,
%``Strangeness production in high-multiplicity events,''
Phys. Rev. D \textbf{104}, no.1, 016024 (2021)




\bibitem{Stebel:2021bbn}
T.~Stebel and K.~Watanabe,
%``J/\ensuremath{\psi} polarization in high multiplicity pp and pA collisions: CGC\,+\,NRQCD approach,''
Phys. Rev. D \textbf{104}, no.3, 034004 (2021)



 



\bibitem{Salazar:2021mpv}
F.~Salazar, B.~Schenke and A.~Soto-Ontoso,
%``Accessing subnuclear fluctuations and saturation with multiplicity dependent J/\ensuremath{\psi} production in p+p and p+Pb collisions,''
Phys. Lett. B \textbf{827}, 136952 (2022)

\bibitem{Lima:2023dqw}
Y.~N.~Lima, A.~V.~Giannini and V.~P.~Goncalves,
%``Isolated photon production in high multiplicity events at the LHC,''
Eur. Phys. J. A \textbf{60}, no.3, 54 (2024)
%0 citations counted in INSPIRE as of 08 Jul 2023




\bibitem{Lima:2022mol}
Y.~N.~Lima, A.~V.~Giannini and V.~P.~Goncalves,
%``Kaon production in high multiplicity events at the Large Hadron Collider,''
Phys. Rev. C \textbf{106}, no.6, 065206 (2022)

\bibitem{Ferreiro:2015gea}
E.~G.~Ferreiro and C.~Pajares,
%``Open charm production in high multiplicity proton-proton events at the LHC,''
[arXiv:1501.03381 [hep-ph]].




\bibitem{Werner:2016nsq}
K.~Werner, B.~Guiot, I.~Karpenko, T.~Pierog and G.~Sophys,
%``Charm production in high multiplicity pp events,''
[arXiv:1602.03414 [nucl-th]].


\bibitem{Tripathy:2022teu}
T.~Tripathy, B.~Naik, R.~Nayak, N.~Behera, B.~K.~Nandi and S.~Dash,
%``Study of multiplicity dependence of heavy flavor production in p-p collisions using rope hadronization mechanism,''
[arXiv:2209.08784 [hep-ex]].



\bibitem{BAL}  I. I. Balitsky,   Nucl. Phys. {\bf  B463}, 99 (1996); Phys. Rev. Lett. {\bf 81}, 2024 (1998); Phys. Rev. D  {\bf 60}, 014020 (1999); I. I. Balitsky,   Phys. Lett. B  {\bf 518}, 235 (2001);   I.I. Balitsky and  A.V. Belitsky, Nucl. Phys. {\bf B629}, 290  (2002). 
 

\bibitem{KOVCHEGOV}  
Y.V. Kovchegov,  Phys. Rev. D {\bf 60},  034008 (1999);  Phys. Rev. D {\bf 61} 074018 (2000). 



\bibitem{Brodsky:2015fna}
S.~J.~Brodsky, A.~Kusina, F.~Lyonnet, I.~Schienbein, H.~Spiesberger and R.~Vogt,
%``A review of the intrinsic heavy quark content of the nucleon,''
Adv. High Energy Phys. \textbf{2015}, 231547 (2015)



\bibitem{Goncalves:2008sw}
V.~P.~Goncalves, F.~S.~Navarra and T. Ullrich,
%``Looking for intrinsic charm in the forward region at BNL RHIC and CERN LHC,''
Nucl. Phys. A \textbf{842}, 59-71 (2010)



\bibitem{Carvalho:2017zge}
F.~Carvalho, A.~V.~Giannini, V.~P.~Goncalves and F.~S.~Navarra,
%``$D$ - meson production at very forward rapidities: estimating the intrinsic charm contribution,''
Phys. Rev. D \textbf{96}, no.9, 094002 (2017)

\bibitem{Giannini:2018utr}
A.~V.~Giannini, V.~P.~Gon\c{c}alves and F.~S.~Navarra,
%``Intrinsic charm contribution to the prompt atmospheric neutrino flux,''
Phys. Rev. D \textbf{98}, no.1, 014012 (2018)



\bibitem{Maciula:2020dxv}
R.~Maciu\l{}a and A.~Szczurek,
%``Intrinsic charm in the nucleon and charm production at large rapidities in collinear, hybrid and k$_{T}$-factorization approaches,''
JHEP \textbf{10}, 135 (2020)

\bibitem{Goncalves:2021yvw}
V.~P.~Goncalves, R.~Maciula and A.~Szczurek,
%``Impact of intrinsic charm amount in the nucleon and saturation effects on the prompt atmospheric $\nu _{\mu }$ flux for IceCube,''
Eur. Phys. J. C \textbf{82}, no.3, 236 (2022)

\bibitem{Maciula:2022lzk}
R.~Maciula and A.~Szczurek,
%``Far-forward production of charm mesons and neutrinos at forward physics facilities at the LHC and the intrinsic charm in the proton,''
Phys. Rev. D \textbf{107}, no.3, 034002 (2023)




\bibitem{LHCb:2021stx}
R.~Aaij \textit{et al.} [LHCb],
%``Study of Z Bosons Produced in Association with Charm in the Forward Region,''
Phys. Rev. Lett. \textbf{128}, no.8, 082001 (2022)



\bibitem{Ball:2022qks}
R.~D.~Ball \textit{et al.} [NNPDF],
%``Evidence for intrinsic charm quarks in the proton,''
Nature \textbf{608}, no.7923, 483-487 (2022)


\bibitem{NNPDF:2023tyk}
R.~D.~Ball \textit{et al.} [NNPDF],
%``The intrinsic charm quark valence distribution of the proton,''
[arXiv:2311.00743 [hep-ph]].


\bibitem{Goncalves:2017chx}
V.~P.~Goncalves, B.~Kopeliovich, J.~Nemchik, R.~Pasechnik and I.~Potashnikova,
%``Heavy flavor production in high-energy $pp$ collisions: color dipole description,''
Phys. Rev. D \textbf{96}, no.1, 014010 (2017)
%%%%[arXiv:1704.04699 [hep-ph]].


\bibitem{SampaiodosSantos:2021tfh}
G.~Sampaio dos Santos, G.~Gil da Silveira and M.~V.~T.~Machado,
%``Charmed meson production based on dipole transverse momentum representation in high energy hadron-hadron collisions available at the LHC,''
Phys. Lett. B \textbf{838}, 137667 (2023)
%%%%[arXiv:2108.01562 [hep-ph]].


\bibitem{Collins:1991ty}
J.~C.~Collins and R.~K.~Ellis,
%``Heavy quark production in very high-energy hadron collisions,''
Nucl. Phys. B \textbf{360}, 3-30 (1991)



\bibitem{Catani:1990eg}
S.~Catani, M.~Ciafaloni and F.~Hautmann,
%``High-energy factorization and small x heavy flavor production,''
Nucl. Phys. B \textbf{366}, 135-188 (1991)


\bibitem{Nikolaev:2005qs}
N.~N.~Nikolaev, W.~Schafer and B.~G.~Zakharov,
%``Nonuniversality aspects of nonlinear k-perpendicular-factorization for hard dijets,''
Phys. Rev. Lett. \textbf{95}, 221803 (2005)


\bibitem{Gelis:2008rw}
F.~Gelis, T.~Lappi and R.~Venugopalan,
%``High energy factorization in nucleus-nucleus collisions,''
Phys. Rev. D \textbf{78}, 054019 (2008)




\bibitem{Kotko:2015ura}
P.~Kotko, K.~Kutak, C.~Marquet, E.~Petreska, S.~Sapeta and A.~van Hameren,
%``Improved TMD factorization for forward dijet production in dilute-dense hadronic collisions,''
JHEP \textbf{09}, 106 (2015)



\bibitem{Raufeisen:2002ka}
J.~Raufeisen and J.~C.~Peng,
%``Relating parton model and color dipole formulation of heavy quark hadroproduction,''
Phys. Rev. D \textbf{67}, 054008 (2003)


\bibitem{Tuchin:2004rb}
K.~Tuchin,
%``Heavy quark production by a quasiclassical color field in proton nucleus collisions,''
Phys. Lett. B \textbf{593}, 66-74 (2004)


\bibitem{Goncalves:2006ch}
V.~P.~Goncalves and M.~V.~T.~Machado,
%``Saturation Physics in Ultra High Energy Cosmic Rays: Heavy Quark Production,''
JHEP \textbf{04}, 028 (2007)


\bibitem{Cazaroto:2011qq}
E.~R.~Cazaroto, V.~P.~Goncalves and F.~S.~Navarra,
%``Heavy quark production at LHC in the color dipole formalism,''
Nucl. Phys. A \textbf{872}, 196-209 (2011)


\bibitem{Tuchin:2012cd}
K.~Tuchin,
%``Beyond the proton collinear factorization in heavy quark production in pA collisions at low x,''
Nucl. Phys. A \textbf{899}, 44-59 (2013)


\bibitem{Fujii:2013yja}
H.~Fujii and K.~Watanabe,
%``Heavy quark pair production in high energy pA collisions: Open heavy flavors,''
Nucl. Phys. A \textbf{920}, 78-93 (2013)


\bibitem{Altinoluk:2015vax}
T.~Altinoluk, N.~Armesto, G.~Beuf, A.~Kovner and M.~Lublinsky,
%``Heavy quarks in proton-nucleus collisions - the hybrid formalism,''
Phys. Rev. D \textbf{93}, no.5, 054049 (2016)

\bibitem{Bhattacharya:2016jce}
A.~Bhattacharya, R.~Enberg, Y.~S.~Jeong, C.~S.~Kim, M.~H.~Reno, I.~Sarcevic and A.~Stasto,
%``Prompt atmospheric neutrino fluxes: perturbative QCD models and nuclear effects,''
JHEP \textbf{11}, 167 (2016)


\bibitem{Pumplin:2007wg}
J.~Pumplin, H.~L.~Lai and W.~K.~Tung,
%``The Charm Parton Content of the Nucleon,''
Phys. Rev. D \textbf{75}, 054029 (2007)


\bibitem{Kniehl:2006mw}
B.~A.~Kniehl and G.~Kramer,
%``Charmed-hadron fragmentation functions from CERN LEP1 revisited,''
Phys. Rev. D \textbf{74}, 037502 (2006)
%%%%%[arXiv:hep-ph/0607306 [hep-ph]].


\bibitem{Dulat:2013hea}
S.~Dulat, T.~J.~Hou, J.~Gao, J.~Huston, J.~Pumplin, C.~Schmidt, D.~Stump and C.~P.~Yuan,
%``Intrinsic Charm Parton Distribution Functions from CTEQ-TEA Global Analysis,''
Phys. Rev. D \textbf{89}, no.7, 073004 (2014)
%%%%[arXiv:1309.0025 [hep-ph]].










\bibitem{Brodsky:1980pb}
S.~J.~Brodsky, P.~Hoyer, C.~Peterson and N.~Sakai,
%``The Intrinsic Charm of the Proton,''
Phys. Lett. B \textbf{93}, 451-455 (1980)



\bibitem{Navarra:1995rq}
F.~S.~Navarra, M.~Nielsen, C.~A.~A.~Nunes and M.~Teixeira,
%``On the intrinsic charm component of the nucleon,''
Phys. Rev. D \textbf{54}, 842-846 (1996)

\bibitem{Paiva:1996dd}
S.~Paiva, M.~Nielsen, F.~S.~Navarra, F.~O.~Duraes and L.~L.~Barz,
%``Virtual meson cloud of the nucleon and intrinsic strangeness and charm,''
Mod. Phys. Lett. A \textbf{13}, 2715-2724 (1998)


\bibitem{Hobbs:2013bia}
T.~J.~Hobbs, J.~T.~Londergan and W.~Melnitchouk,
%``Phenomenology of nonperturbative charm in the nucleon,''
Phys. Rev. D \textbf{89}, no.7, 074008 (2014)



\bibitem{Golec-Biernat:1998zce}
K.~J.~Golec-Biernat and M.~Wusthoff,
%``Saturation effects in deep inelastic scattering at low Q**2 and its implications on diffraction,''
Phys. Rev. D \textbf{59}, 014017 (1998)
%%%%[arXiv:hep-ph/9807513 [hep-ph]].


\bibitem{McLerran:1997fk}
L.~D.~McLerran and R.~Venugopalan,
%``Boost covariant gluon distributions in large nuclei,''
Phys. Lett. B \textbf{424}, 15-24 (1998)
%%%[arXiv:nucl-th/9705055 [nucl-th]].


\bibitem{Albacete:2009fh}
J.~L.~Albacete, N.~Armesto, J.~G.~Milhano and C.~A.~Salgado,
%``Non-linear QCD meets data: A Global analysis of lepton-proton scattering with running coupling BK evolution,''
Phys. Rev. D \textbf{80}, 034031 (2009)
%doi:10.1103/PhysRevD.80.034031
%%%%[arXiv:0902.1112 [hep-ph]].


\bibitem{Albacete:2010sy}
J.~L.~Albacete, N.~Armesto, J.~G.~Milhano, P.~Quiroga-Arias and C.~A.~Salgado,
%``AAMQS: A non-linear QCD analysis of new HERA data at small-x including heavy quarks,''
Eur. Phys. J. C \textbf{71}, 1705 (2011)
%%%%[arXiv:1012.4408 [hep-ph]].




\bibitem{Watanabe:2015yca}
K.~Watanabe and B.~W.~Xiao,
%``Forward Heavy Quarkonium Productions at the LHC,''
Phys. Rev. D \textbf{92}, no.11, 111502 (2015)
%%%% [arXiv:1507.06564 [hep-ph]].


\bibitem{Benic:2022ixp}
S.~Beni\'c, O.~Garcia-Montero and A.~Perkov,
%``Isolated photon-hadron production in high energy pp and pA collisions at RHIC and LHC,''
Phys. Rev. D \textbf{105}, no.11, 114052 (2022)
%%%%[arXiv:2203.01685 [hep-ph]].



\bibitem{Bhattacharya:2023zei}
A.~Bhattacharya, F.~Kling, I.~Sarcevic and A.~M.~Stasto,
%``Forward Neutrinos from Charm at Large Hadron Collider,''
Phys. Rev. D \textbf{109}, no.1, 014040 (2024)

\bibitem{Guiot:2021vnp}
B.~Guiot and A.~van Hameren,
%``D and B-meson production using kt-factorization calculations in a variable-flavor-number scheme,''
Phys. Rev. D \textbf{104}, no.9, 094038 (2021)
%%%%%[arXiv:2108.06419 [hep-ph]].

\bibitem{Xie:2021ycd}
K.~Xie, J.~M.~Campbell and P.~M.~Nadolsky,
%``A general-mass scheme for prompt charm production at hadron colliders,''
SciPost Phys. Proc. \textbf{8}, 084 (2022)
%%%%[arXiv:2108.03741 [hep-ph]].


\bibitem{Bai:2021ira}
W.~Bai, M.~Diwan, M.~V.~Garzelli, Y.~S.~Jeong, F.~K.~Kumar and M.~H.~Reno,
%``Parton distribution function uncertainties in theoretical predictions for far-forward tau neutrinos at the Large Hadron Collider,''
JHEP \textbf{06}, 148 (2022)
%%%[arXiv:2112.11605 [hep-ph]].

\bibitem{Santos:2022eee}
G.~S.~d.~Santos, G.~G.~da Silveira and M.~V.~T.~Machado,
%``D-meson production in high energy pA collisions within the QCD color dipole transverse momentum representation,''
Eur. Phys. J. C \textbf{82}, no.9, 795 (2022)
%%%%[arXiv:2205.00925 [hep-ph]].
%2 citations counted in INSPIRE as of 07 Jul 2023




%\bibitem{Guzzi:2022rca}
%M.~Guzzi, T.~J.~Hobbs, K.~Xie, J.~Huston, P.~Nadolsky and C.~P.~Yuan,
%``The persistent nonperturbative charm enigma,''
%Phys. Lett. B \textbf{843}, 137975 (2023)


\bibitem{AbdulKhalek:2021gbh}
R.~Abdul Khalek, A.~Accardi, J.~Adam, D.~Adamiak, W.~Akers, M.~Albaladejo, A.~Al-bataineh, M.~G.~Alexeev, F.~Ameli and P.~Antonioli, \textit{et al.}
%``Science Requirements and Detector Concepts for the Electron-Ion Collider: EIC Yellow Report,''
Nucl. Phys. A \textbf{1026}, 122447 (2022)

\bibitem{Burkert:2022hjz}
V.~D.~Burkert, L.~Elouadrhiri, A.~Afanasev, J.~Arrington, M.~Contalbrigo, W.~Cosyn, A.~Deshpande, D.~I.~Glazier, X.~Ji and S.~Liuti, \textit{et al.}
%``Precision studies of QCD in the low energy domain of the EIC,''
Prog. Part. Nucl. Phys. \textbf{131}, 104032 (2023)

\bibitem{Anchordoqui:2021ghd}
L.~A.~Anchordoqui, A.~Ariga, T.~Ariga, W.~Bai, K.~Balazs, B.~Batell, J.~Boyd, J.~Bramante, M.~Campanelli and A.~Carmona, \textit{et al.}
%``The Forward Physics Facility: Sites, experiments, and physics potential,''
Phys. Rept. \textbf{968}, 1-50 (2022)



\bibitem{Feng:2022inv}
J.~L.~Feng, F.~Kling, M.~H.~Reno, J.~Rojo, D.~Soldin, L.~A.~Anchordoqui, J.~Boyd, A.~Ismail, L.~Harland-Lang and K.~J.~Kelly, \textit{et al.}
%``The Forward Physics Facility at the High-Luminosity LHC,''
J. Phys. G \textbf{50}, no.3, 030501 (2023)





\bibitem{ALICE:2015ikl}
J.~Adam \textit{et al.} [ALICE],
%``Measurement of charm and beauty production at central rapidity versus charged-particle multiplicity in proton-proton collisions at $ \sqrt{s}=7 $ TeV,''
JHEP \textbf{09}, 148 (2015)


\bibitem{ALICE:2017wet}
D.~Adamov\'a \textit{et al.} [ALICE],
%``J/$\psi$ production as a function of charged-particle pseudorapidity density in p-Pb collisions at $\sqrt{s_{\rm NN}} = 5.02$ TeV,''
Phys. Lett. B \textbf{776}, 91-104 (2018)

\bibitem{STAR:2018smh}
J.~Adam \textit{et al.} [STAR],
%``$J/\psi$ production cross section and its dependence on charged-particle multiplicity in $p + p$ collisions at $\sqrt{s}$ = 200 GeV,''
Phys. Lett. B \textbf{786}, 87-93 (2018)


\bibitem{ALICE:2020msa}
S.~Acharya \textit{et al.} [ALICE],
%``Multiplicity dependence of J/$\psi$ production at midrapidity in pp collisions at $\sqrt{s}$ = 13 TeV,''
Phys. Lett. B \textbf{810}, 135758 (2020)

\bibitem{ALICE:2020eji}
S.~Acharya \textit{et al.} [ALICE],
%``J/$\psi$ production as a function of charged-particle multiplicity in p-Pb collisions at $\sqrt{\textit{s}_{\rm NN}}~=~8.16$ TeV,''
JHEP \textbf{09}, 162 (2020)





\bibitem{ALICECollaboration:2020}
ALICE Collaboration,
%\emph{Multiplicity dependence of (multi-)strange hadron production in proton-proton collisions at $\sqrt{s} = 13$ TeV},
%\href{https://doi.org/10.1140/epjc/s10052-020-7673-8}
Eur. Phys. J. C \textbf{80}, 167 (2020).


\bibitem{ALICE:2021zkd}
S.~Acharya \textit{et al.} [ALICE],
%``Forward rapidity J/\ensuremath{\psi} production as a function of charged-particle multiplicity in pp collisions at $ \sqrt{s} $ = 5.02 and 13 TeV,''
JHEP \textbf{06}, 015 (2022)


\bibitem{Blaizot:2004wv}
J.~P.~Blaizot, F.~Gelis and R.~Venugopalan,
%``High-energy pA collisions in the color glass condensate approach. 2. Quark production,''
Nucl. Phys. A \textbf{743}, 57-91 (2004)



%%%%%%%%%%%%%%%%%%%%%%%%%%%%%%%%%%%%%%%



\bibitem{Alvioli:2012ba}
M.~Alvioli, G.~Soyez and D.~N.~Triantafyllopoulos,
%``Testing the Gaussian Approximation to the JIMWLK Equation,''
Phys. Rev. D \textbf{87}, no.1, 014016 (2013)


\bibitem{Marquet:2016cgx}
C.~Marquet, E.~Petreska and C.~Roiesnel,
%``Transverse-momentum-dependent gluon distributions from JIMWLK evolution,''
JHEP \textbf{10}, 065 (2016)




\bibitem{Shi:2017gcq}
Y.~Shi, C.~Zhang and E.~Wang,
%``Multipole scattering amplitudes in the Color Glass Condensate formalism,''
Phys. Rev. D \textbf{95}, no.11, 116014 (2017)

\bibitem{Fukushima:2017mko}
K.~Fukushima and Y.~Hidaka,
%``General formulae for dipole Wilson line correlators with the Color Glass Condensate,''
JHEP \textbf{11}, 114 (2017)


\bibitem{Banu:2019uei}
K.~Banu, M.~Siddiqah and R.~Abir,
%``Small-$x$ evolution of 2$n$-tuple Wilson line correlator revisited: The nonsingular kernels,''
Phys. Rev. D \textbf{99}, no.9, 094017 (2019)





\end{thebibliography}
\end{document}